\documentclass[mathleft
]{an}
\usepackage{graphicx}
\usepackage{times}
\begin{document}

% The following seven commands are intended for editorial usage and should be ignored by
% the author(s).
\Pagespan{789}{}% Document's page range. 
% If second parameter is left empty, the last page is computed automatically.
\Yearpublication{2006}%
\Yearsubmission{2005}%
\Month{11}%   
\Volume{999}%  
\Issue{88}% 
% \DOI{This.is/not.aDOI}% 

\title{X-ray observations and the search for {\sl Fermi}-LAT $\gamma$-ray pulsars}

\author{P. M. Saz Parkinson\inst{1}\inst{2}
, A. Belfiore\inst{1}, P. Caraveo\inst{3}, A. De Luca\inst{3}, and M. Marelli\inst{3} 
for the {\it Fermi} LAT Collaboration \newline \fnmsep\email{pablo@scipp.ucsc.edu}}
\titlerunning{X-rays and LAT pulsars}
\authorrunning{P. M. Saz Parkinson}
\institute{Santa Cruz Institute for Particle Physics, University of California, Santa Cruz, CA 95064
\and 
Department of Physics, The University of Hong Kong, Pokfulam Road, Hong Kong
\and
INAF -- Istituto di Astrofisica Spaziale e Fisica Cosmica, Via Bassini 15, 20133 Milano, Italy
}

\received{26 October 2013}
\accepted{xxx}
\publonline{xxx}


\abstract{%
The Large Area Telescope (LAT) on {\it Fermi} has detected $\sim$150 $\gamma$-ray pulsars, about a third of
which were discovered in blind searches of the $\gamma$-ray data. Because
the angular resolution of the LAT is relatively poor and blind
searches for pulsars (especially millisecond pulsars, MSPs) are
very sensitive to an error in the position, one must typically scan large
numbers of locations. Identifying plausible X-ray counterparts of a
putative pulsar drastically reduces the
number of trials, thus improving the sensitivity of pulsar blind
searches with the LAT. I discuss our ongoing program of
{\it Swift}, XMM-{\it Newton}, and {\it Chandra} observations of LAT
unassociated sources in the context of our blind searches for
$\gamma$-ray pulsars.}
\maketitle

\section{Introduction}

\begin{figure}
\includegraphics[width=80mm]{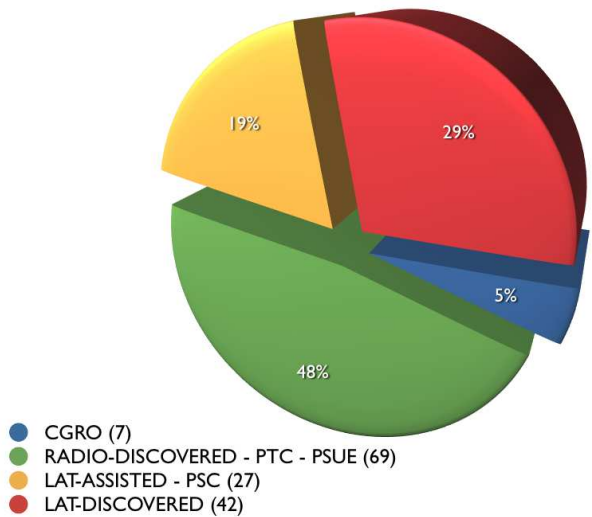}
\caption{{\bf The $\gamma$-ray pulsar pie} -- Distribution of the 145 $\gamma$-ray
  pulsars in the LAT 2nd Pulsar Catalog (2PC, [\cite{2PC}]). In blue are {\it Compton Gamma-ray Observatory} (CGRO)
  $\gamma$-ray pulsars, known before the launch of {\it
    Fermi}. In green are $\gamma$-ray pulsars detected using radio
  ephemerides. In yellow are those discovered in radio searches
  of LAT unassociated sources, subsequently shown to pulse in $\gamma$
  rays (the so-called {\it LAT-assisted} $\gamma$-ray
  pulsars). In red are pulsars discovered in {\it blind searches} of LAT data.}
\label{fig1}
\end{figure}

Since its launch, in June of 2008, the {\it Fermi} Large Area Telescope (LAT) has dramatically increased the
number of known $\gamma$-ray pulsars. At the end of the {\it Compton Gamma Ray Observatory (CGRO)} era there were 7 {\it firm}
detections~[\cite{Thompson04}], while the Second  $Fermi$ LAT Catalog
of Gamma-ray Pulsars (2PC) contains 145 $\gamma$-ray pulsars detected in
the first 3 years of the {\it Fermi} mission~[\cite{2PC}]. Almost a third (42) of the 2PC $\gamma$-ray
pulsars were discovered in blind searches of LAT data
(e.g. [\cite{BSI,BSII,BSIII}], See Figure~\ref{fig1}). Radio follow-up
observations of the these LAT-discovered pulsars failed to detect pulsations for all
but a few of them (e.g. [\cite{camilo09}]), which suggests that a majority of the underlying
population of young, isolated pulsars is, in fact, {\it radio
  quiet}. Thus, X-ray observations may prove to be the only
alternative channel to learn about such pulsars, or even to identify
them, prior to their discovery by the LAT.

\subsection{LAT searches for pulsars}

Blind searches for pulsars in LAT data are complicated by a number of
factors, including the scarcity of photons~\footnote{Typically only a few
photons per day or, equivalently, one photon per thousands or even millions of rotations of the pulsar.} and the relatively poor
point spread function (PSF) of the instrument, as well as its strong
dependence on energy~\footnote{The LAT PSF varies from several degrees at 100 MeV to a few arc minutes at 100 GeV.}. 
X-ray observations of LAT $\gamma$-ray sources significantly improve
the sensitivity of pulsar blind searches by providing 
plausible counterparts with precise ($\sim$arc second) positions~[\cite{Dormody}]. 

\begin{figure}
\includegraphics[width=80mm]{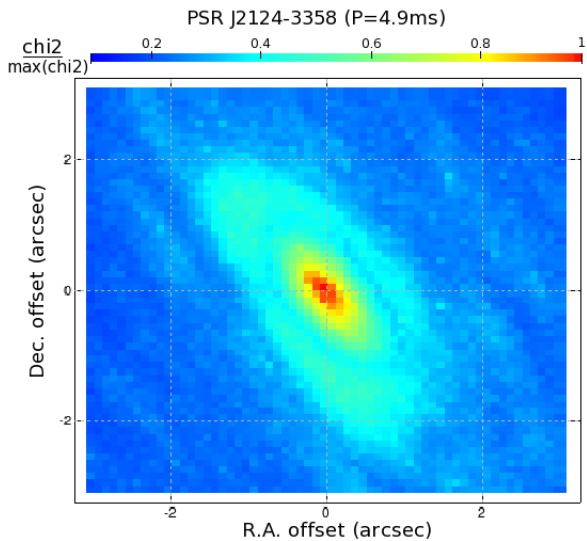}
\caption{Relative significance of the detection of pulsations from the isolated MSP~J2124--3358. The normalized $\chi^2$ is shown as a function of the offset from the radio position. The orientation of the ellipse follows ecliptic coordinates and the area satisfying $\frac{\chi^2}{\chi^2_{max}}>0.5$ is approximately one square arc second, which illustrates the required accuracy for such searches.}
\label{fig2}
\end{figure}

Blind searches for {\it even} isolated $\gamma$-ray pulsars are computationally
intensive, as they involve a search over the fairly large
parameter space of position, frequency, and frequency
derivative. Searches for MSPs, in particular, are extremely sensitive
to position. Typically, one must pick a position and {\it barycenter}
the $\gamma$-ray photons to within 1'', or better, of the
true position of the (undiscovered) pulsar. Figure~\ref{fig2} demonstrates this
sensitivity to the position by showing the drop in relative significance of
the pulsation from the known MSP J2124--3358, when the offset in
position is greater than a small fraction of an arc second. If the $\gamma$-ray source is an
MSP, and we are able to detect {\it a priori} its X-ray counterpart,
then our starting point is generally within at most a few arc seconds
of the true pulsar position. Since typical uncertainties in the LAT
positions are of order arc minutes, the gain in computer time achieved
by scanning the X-ray (as opposed to the $\gamma$-ray) error region is of
order $10^3$. Aside from this gain in computational efficiency,
however, restricting our blind searches to known X-ray positions has
the added advantages of reducing the number of trials,
thus improving the sensitivity of our searches.

The discovery of $\gamma$-ray pulsations in a LAT unassociated source
first and foremost identifies the nature of the source. Its subsequent
study, however, has numerous benefits: the timing analysis can lead to
localizations of unprecedented (for $\gamma$-ray astronomy) accuracy,
rivaling those of X-ray instruments, as described in
[\cite{ray11}]. Follow-up X-ray observations can then unambiguously
identify a counterpart and its further study can tell us more about
the nature of the compact object (e.g. its distance) or the pulsar
population as a whole [\cite{Marelli11}]. Further studies in $\gamma$
rays (e.g. phase-resolved spectroscopy), in conjunction with
theoretical modeling might also reveal further insights into some of
the ongoing questions in pulsar physics (e.g. see Venter \& Harding, in
these proceedings).

In this paper, we describe our ongoing efforts to use {\it Swift},
{\it Chandra}, and {\it XMM-Newton} observations to
enhance our blind searches for pulsars with the LAT.

\section{The multi-wavelength nature of pulsars}

Pulsars are thought to be rapidly-spinning highly-magnetized neutron stars. 
Although the vast majority of the $\sim$2000 known pulsars (including
the first discovered) were found in radio, pulsars are the
quintessential multi-wavelength (MWL) objects, 
having been  detected across the entire electromagnetic spectrum, from the
optical, to the X-ray and $\gamma$-ray bands
[e.g. \cite{Thompson04}]. While optical pulsations are rarely
detected [e.g. see Section 9, 2PC], there are currently over 100 known X-ray pulsars
[\cite{Becker09}]. The nature of the X-ray emission can be
varied, and can include thermal emission from the surface of the
neutron star as well as magnetospheric emission. Even when optical
and/or X-ray pulsations from a pulsar are not detected (often simply due
to the faintness of such emission), it is fairly common to detect an
optical or X-ray counterpart source at the location of the pulsar and
the nature of such a source can lead to its identification as a neutron star (or at least a neutron star
{\it candidate}). 

\begin{figure}
\includegraphics[width=80mm]{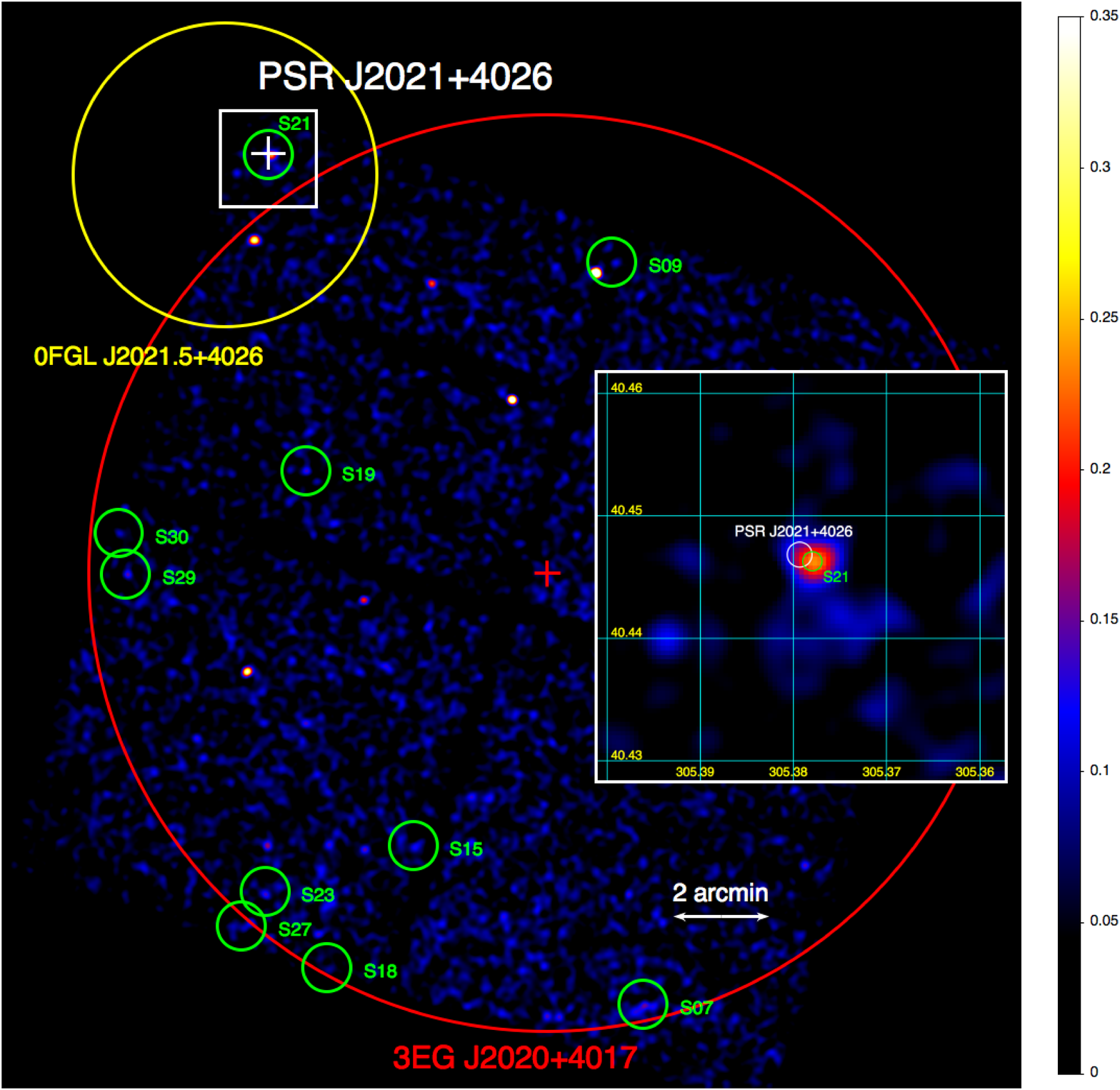}
\caption{{\it Chandra} 14.3 ks ACIS-I observation of the
  unidentified EGRET source 3EG J2020+4017 (PI: Weisskopf). The largest (red) circle
  represents the uncertainty in the EGRET position, while the 
smaller (yellow) circle in the top left represents
    the {\it Fermi}-LAT error circle with just 3 months of data
    (i.e. 0FGL). The numbering of the X-ray sources follows Weisskopf et al. 2006, with
    circled sources representing candidate X-ray
    counterparts. The one surrounded by a box (S21) represents the
    actual X-ray counterpart, whose X-ray position was used in the
    original discovery of the pulsar [\cite{BSI}]. The zoomed-in image shows the error ellipse obtained from
    timing the pulsar in $\gamma$ rays, in good agreement with the X-ray
    counterpart position.}
\label{fig3}
\end{figure}

The X-ray/$\gamma$-ray connection in the studies of pulsars has a long
history. The first known {\it radio-quiet} pulsar, {\it Geminga}, was
first detected as a bright unidentified $\gamma$-ray source by SAS-2
in 1972 but its true nature was only revealed 20 years later through
the detection of pulsations in the X-ray band [\cite{halpern92}], later
confirmed to be present in the $\gamma$-ray band. Although subsequent
analyses have shown that it could have actually been
discovered directly in blind searches of the $\gamma$-ray
data\footnote{Possibly going as far back as the COS-B era.} [e.g. \cite{ziegler}], undoubtedly it was the X-ray studies of
this pulsar that led the way in improving our understanding of this
enigmatic source [for a detailed review of Geminga, see \cite{geminga}]. 
More recently, we have an example of a different path:
PSR~J2021+4026 is a radio-quiet pulsar discovered in early blind searches
of LAT $\gamma$-ray data [\cite{BSI}]. In this case, {\it Chandra}
observations of a bright EGRET unidentified source led to a number of
{\it plausible} candidates [See Figure~\ref{fig3}, adapted from \cite{weisskopf06}]. Following the discovery of $\gamma$-ray
pulsations, further {\it Chandra} observations, in conjuntion with the
pulsar timing performed with the $\gamma$-ray data led to the
identification of its X-ray counterpart [\cite{weisskopf11}], while a deep {\it XMM}
observation of the X-ray counterpart managed to finally uncover X-ray pulsations [\cite{lin13}]. 

Pulsars (especially ``young'' ones) are naturally associated with
supernova remnants (SNRs) and Pulsar Wind Nebulae (PWNe), both of
which are important sources of radiation (including X-rays) in their
own right. Hence, despite neutron stars often being inherently faint
X-ray objects, pulsars can often reveal themselves through these other
sources of unpulsed (most often extended) X-ray emission. 

In short, although X-ray fluxes from most pulsars are extremely low, especially when
compared to their $\gamma$-ray flux [\cite{Marelli11}], the X-ray
window remains one of the most fruitful channels for the discovery and
subsequent study of pulsars. With the launch of {\it Fermi}, X-ray
observations have become, if anything, even more important, as the
population of {\it radio-quiet} neutron stars has vastly increased,
along with our desire to understand these systems in the broader
context of the overall neutron star population.

\section{X-ray observations of LAT unassociated sources}

From the beginning of $\gamma$-ray astronomy, a large number of
$\gamma$-ray sources have been discovered with no known
counterparts (e.g. Geminga). This is a natural consequence of the poor angular
resolution of $\gamma$-ray instruments, relative to those in most
other wavelengths. The first COS-B catalog, for example, contained 25
sources, of which $>$80\% were unidentified. The Third EGRET Catalog 
contained $\sim$300 sources with more than 50\% classified as 
``unidentified''. With the advent of the LAT, the number of $\gamma$-ray
sources has increased dramatically. Although still numbering in the
hundreds, the fraction of sources considered {\it unassociated}\footnote{LAT catalogs distinguish
  between {\it identification} and {\it association} of a $\gamma$-ray
  source with a known astrophysical object, with the former
  requiring a measurement of correlated variability at other
  wavelengths.} continues to fall: from 40\% in the Bright
Source List [0FGL, \cite{0FGL}], to 30\% in the 1FGL catalog [\cite{1FGL}], and $\sim$25\% in the
2FGL catalog [\cite{2FGL}]. This is in large part due to the significant improvement
in the instrument characteristics of the LAT relative to past
missions, but is also the result of the vast multi-wavelength
campaigns that have been mounted to try to identify these sources, in
particular in the radio [e.g. \cite{PSC}] and X-ray bands
[e.g. \cite{Marelli11}]. Figure~\ref{fig3} illustrates the dramatic
improvement in source localization obtained with the LAT, relative to
EGRET, which greatly facilitates multi-wavelength (particularly radio
and X-ray) follow-up observations of these sources.

The nature of LAT unassociated sources can be investigated, to a
certain extent, by comparing their properties to those of $\gamma$-ray
sources in known classes. In particular, the two largest classes of
$\gamma$-ray sources are blazars and pulsars. Fortunately, these
sources differ greatly in their timing and spectral
properties. Blazars are highly variable on long ($>$1 day) time scales
and have power law spectra, while pulsars are fairly steady (although,
see [\cite{J2021}]) and have spectra that are better fit with a power law with an
exponential cutoff. A number of techniques can be used to
statistically classify sources into {\it
  pulsar candidates} or {\it blazar candidates}, by virtue of these
spectral and variability characteristics [e.g.\cite{unassociated}].

\subsection{{\it Swift} follow-up observations of LAT unassociated sources}

\begin{figure}
\includegraphics[width=80mm]{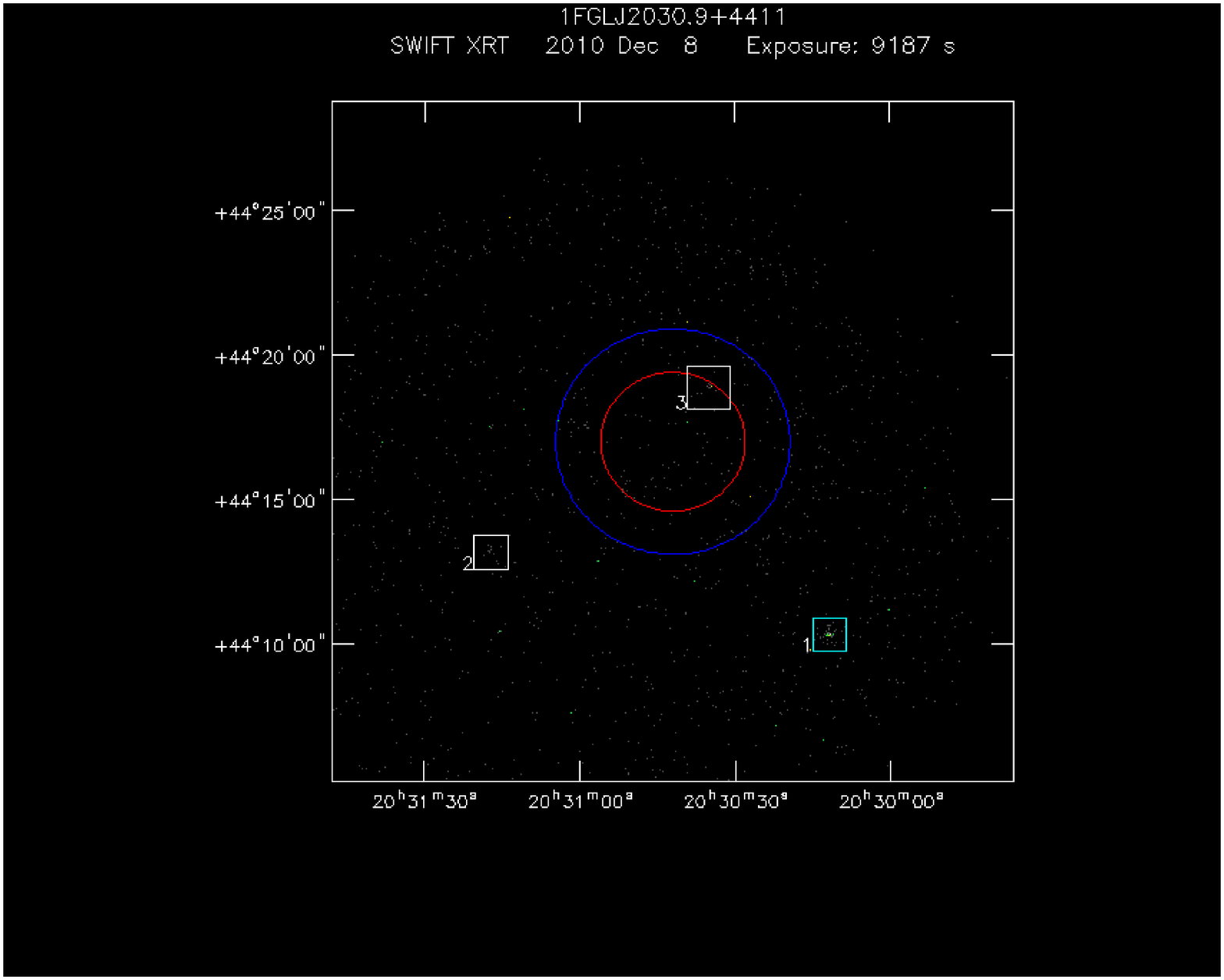}
\caption{{\it Swift} 9.2 ks observation of the LAT {\it pulsar-like} unassociated source
  1FGL~J2030.9+441, showing a number of plausible X-ray counterparts. For more details, go to {\tt
  http://www.swift.psu.edu/unassociated/}. }
\label{fig4}
\end{figure}

As part of a successful {\it Fermi} Guest Investigator Cycle 3 (later
renewed in Cycles 4 and 5) program~[\cite{swift}], we
organized a campaign of follow-up observations of {\it Fermi}-LAT
unassociated sources with {\it Swift}. Starting with the unassociated
sources in the 1FGL catalog, our program was later expanded to include unassociated
sources in 2FGL. Table~\ref{swift_table} gives a summary of the {\it
  Swift} observations carried out to date, as part of this program,
along with the number of X-ray sources detected. The goals of this shallow survey are simply to
identify plausible X-ray counterparts of the LAT unassociated
sources. We previously determined which of these sources might be
promising pulsar candidates (as described in the previous section),
and typically observed these LAT sources for $\sim$10 ks, while the
remaining unassociated sources (overwhelmingly blazar candidates,
which generally have brighter X-ray counterparts) received only short ($\sim$4 ks)
observations. Figure~\ref{fig4} shows the results of a 9.2 ks {\it Swift}
observation of the LAT {\it pulsar-like} unassociated source 1FGL~J2030.9+441.
The results of all the {\it Swift} observations are summarized in the
publicly available web site: {\tt http://www.swift.psu.edu/unassociated/}. 
We then proceeded to select all X-ray counterparts
of the {\it pulsar candidates} and perform blind searches of the
$\gamma$-ray data using the precise {\it Swift} locations. The LAT
blind searches on these X-ray counterpart locations are still in
progress and the results of such searches will be reported in a future publication.

\begin{table}
% \centering%%%
\caption{Swift X-ray observations of LAT {\it pulsar-like}
  unassociated sources (As of August 2013).}
\label{swift_table}
\begin{tabular}{cccc}\hline
Catalog & \# Sources (Unass.)&{\it Swift} Obs. & X-ray Sources  \\
                    &                                          & &(SNR$>$3)   \\
\hline
1FGL &  1451 (630) &  251 & 544\\
2FGL &  1873 (575) &  179 & 198\\

\hline
\end{tabular}
\end{table}

\subsection{XMM-{\it Newton} Observations}

In addition to our {\it Swift} campaign, we proposed deeper
observations with more sensitive X-ray missions such as {\it Chandra}
and XMM-{\it Newton} of a few `select' candidates. Our XMM campaign focused on trying to identify
X-ray counterparts of three bright high-Galactic-latitude LAT
$\gamma$-ray sources which were plausible candidate radio-quiet MSPs
(See Table~\ref{xmm_table}). These targets were chosen as being
extremely promising pulsar candidates (by some of the statistical
measures alluded to in the previous section), apart from being relatively bright in $\gamma$
rays. It was thus felt that they were {\it deserving} of the better angular and spectral
characteristics of {\it XMM}. Indeed, one of our three targets (J1536.4--4949)
subsequently turned out to host an MSP, discovered in a radio search of
LAT unassociated sources carried out by the GMRT\footnote{{\tt http://gmrt.ncra.tifr.res.in}}[\cite{PSC}]. 

\begin{table}
% \centering%%%
\caption{XMM X-ray observations of {\it pulsar-like} 2FGL unassociated sources}
\label{xmm_table}
\begin{tabular}{ccc}\hline
Target &  duration &  \\ 
(2FGL) & (ks) &  \\
\hline
J1744.1--7620 & 26 &  \\
J1036.1--6722 &  25 & \\
J1536.4--4949 &  18 & \\

\hline
\end{tabular}
\end{table}

\subsection{{\it Chandra} Observations}

The {\it Fermi} LAT has detected several dozen Supernova Remnants
(SNRs) and many of the newly-discovered $\gamma$-ray pulsars are
positionally coincident (and possibly associated) with SNRs. 
Indeed, the number of pulsar-SNR associations has grown dramatically
thanks to {\it Fermi}. Disentangling the $\gamma$-ray emission due to
the pulsar from that due to the SNR is crucial to explaining the
nature of such emission. 

{\it Chandra} can be extremely useful in identifying potential pulsar candidates in bright
$\gamma$-ray sources with pulsar-like properties, and those which are
already known to be associated with SNRs are therefore prime
candidates to be targeted. In addition, the excellent angular resolution of {\it Chandra} raises the possibility of detecting extended
X-ray emission from a PWN, making such a detection a strong indication
for the presence of a young pulsar. 

\begin{figure}
\begin{center}
\includegraphics[width=3in]{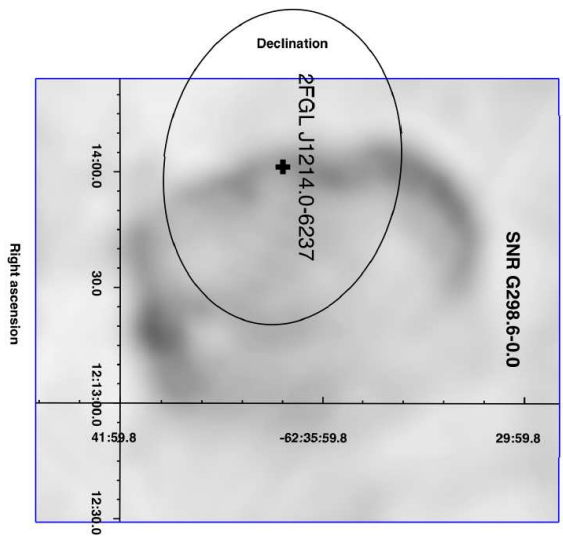}
\end{center}
\caption{
843 MHz Radio continuum image showing the extent of SNR G298.6-0.0. The location of the nearby 2FGL source is shown with an error ellipse tracing out the 95\% localization boundary. The ACIS-I field of view covers the entire image, including the 2FGL source.
}
\label{fig5}
\end{figure}

With this goal in mind, we proposed (and obtained) a number of {\it Chandra}
observations of such sources, in the hopes of discovering the
(probably radio-quiet) $\gamma$-ray pulsar associated with such SNRs
(see Table~\ref{chandra_table}). Figure~\ref{fig5}, for example, shows the
positional coincidence of the LAT source 2FGL~J1214.0--6237 and SNR
G298.6--0.0. The analysis of these X-ray observations is still in progress.

\begin{table}
% \centering%%%
\caption{Chandra X-ray observations of LAT-detected SNRs with {\it
    pulsar-like} $\gamma$-ray emission}
\label{chandra_table}
\begin{tabular}{ccc}\hline
SNR &  duration &  \\ 
(2FGL) & (ks) &  \\
\hline
G298.6--0.0 (2FGL~J1214.0--6237) & 20 &  \\
G311.50.3 (2FGL~J1405.5--6121) &  13 & \\
G359.1+0.9 (2FGL~J1738.9--2908) &  11 & \\

\hline
\end{tabular}
\end{table}

\section{Searches for {\it Black Widows}}

Unfortunately, a majority of MSPs are in binary systems, adding an
extra complication to the blind search for such pulsars. Indeed, over 85\% of MSPs discovered in LAT sources fall in
this category. In fact, a large number of these are in so-called {\it
  Black-Widow} (or Redback) systems. These systems have extremely
tight circular orbits and are often subject to wide eclipses, making
them quite challenging to detect in radio. Although no clear orbital modulation has been
detected in the $\gamma$-ray light curves of these objects, optical
(and sometimes X-ray) observations can clearly reveal the binary
nature of the system. 

Despite the fact that {\it Black Widow}-like binaries have almost
circular orbits, thus requiring a search over ``only'' three additional
orbital parameters, a full blind search of the $\gamma$-ray data would
be unfeasible with current available computer power. However, an
analysis of the X-ray and optical light curves, along with optical
spectral observations and some modeling, can result in fairly robust
estimates for at least two of the three additional parameters [e.g. \cite{romani13}], thus making it possible to perform a {\it
  semi-blind} search of the LAT data. Such a search has already proven
successful~[\cite{J1311}], and more such searches are planned for the future.

\acknowledgements
The $Fermi$ LAT Collaboration acknowledges support from a number of
agencies and institutes for both development and the operation of the
LAT as well as scientific data analysis. These include NASA and DOE in
the United States, CEA/Irfu and IN2P3/CNRS in France, ASI and INFN in
Italy, MEXT, KEK, and JAXA in Japan, and the K.~A.~Wallenberg
Foundation, the Swedish Research Council and the National Space Board
in Sweden. Additional support from INAF in Italy and CNES in France
for science analysis during the operations phase is also gratefully
acknowledged. This work was supported by the {\it Fermi} GI Program (Cycles 3, 4, and 5) 
under NASA grants NNX10AP18G, NNX12AP41G, and NNG12PP66P as well as 
by the XMM-Newton AO-11 cycle under NASA Grant NNX13AB52G. We would like to thank Neil Gehrels for his
continued support of our {\it Swift} folow-up program
and Abe Falcone, and the rest of the {\it Swift} team for leading and
implementing it. Finally, I thank the organizers of the workshop for their hospitality
and for making this a very successful and enjoyable meeting.

\newpage%%%%%%%%%%%%%%%%%%%%%%%%%%%%%%%%%%%%%%%%%%%%%%%%%%%%%%

\end{document}